\documentclass[aps,prd]{revtex4}
\usepackage{amsmath,amssymb}

\newcommand{\be}{\begin{equation}}
\newcommand{\ee}{\end{equation}}
\newcommand{\ba}{\begin{eqnarray}}
\newcommand{\ea}{\end{eqnarray}}
\newcommand{\ban}{\begin{eqnarray*}}
\newcommand{\ean}{\end{eqnarray*}}
\newcommand \nn {\nonumber}

\begin{document}

\title{Wigner functional of fermionic fields}

\author{Stanis\l aw Mr\' owczy\' nski}

\affiliation{Institute of Physics, Jan Kochanowski University, Kielce, Poland}
\affiliation{National Centre for Nuclear Research, Warsaw, Poland}

\date{April 3, 2013}

\begin{abstract}

The Wigner function, which provides a phase-space description of quantum systems, has various applications in quantum mechanics, quantum kinetic theory, quantum optics, radiation transport  and others. The concept of Wigner function has been extended to quantum fields, scalar and electromagnetic. Then, one deals with the Wigner {\em functional} which gives a distribution of field and its conjugate momentum. We introduce here the Wigner functional of fermionic fields of the values in a Grassmann algebra. Properties of the functional are discussed and its equation of motion, which is of the Liouville's form, is derived. 

\end{abstract}



\maketitle

\section{Introduction}

A description of a quantum mechanical system is usually formulated in a position or momentum space. A probabilistic description in a phase space spanned by positions and momenta is not possible because of the uncertainty principle. The quasiprobabilistic approach can be achieved by using the Wigner function \cite{Wigner:1932}, which for a system of single degree of freedom, is usually defined nowadays as
\be
\label{Wigner-function-def}
W(x,p;t) \buildrel \rm def \over = \int du \, e^{-ipu}  \langle x +u/2| \hat\rho(t) |x -u/2 \rangle,
\ee
where $\hat\rho(t)$ is the time-dependent density operator and $ |x \rangle$ is the position eigenstate. The Wigner function is real but it is not everywhere positive, and consequently the approach is not probabilistic but only  quasiprobabilistic.  Various quantum mechanical problems can be formulated in terms of the Wigner function \cite{Carruthers:1982fa}.  It is particularly useful to discuss the classical limit of quantum mechanics. In case of many-body systems, the Wigner function plays a role of quantum analog of classical distribution function, see {\it e.g.} \cite{Kadanoff-Baym:1962}, and it is thus the key object of quantum kinetic theory.  The phase-space methods are particularly well developed in the theory of radiation transport and quantum optics, see the monographs \cite{Apresyan-Kravtsov:1996,Schleich:2001}.

The concept of the Wigner function was extended to quantum fields in \cite{Mrowczynski:1994nf} where a real scalar field  was discussed. The Wigner {\em functional} is defined in a full analogy to Eq.~(\ref{Wigner-function-def}) that is 
\ba
\label{WF-Bose-def}
W[\Phi,\Pi;t] \buildrel \rm def \over = 
\int \mathcal{D}\varphi({\bf x})\, \mathrm{exp}\Big[-i\int d^3x\ \Pi({\bf x})\varphi({\bf x})\Big]
\langle \Phi({\bf x}) + \frac{1}{2}\varphi({\bf x})|\hat{\rho}(t)|\Phi({\bf x}) - \frac{1}{2}\varphi({\bf x})\rangle,
\ea
where $\Phi({\bf x})$ is the time-independent scalar field in the Schr\" odinger picture, ${\bf x} \in \mathbb{R}^3$ is the coordinate variable and $\int \mathcal{D}\varphi({\bf x}) \dots$ denotes the functional integral. The Wigner functional was shown to be a density in an infinitely dimensional phase space spanned by $\Phi({\bf x})$ and its conjugate momentum $\Pi({\bf x})$. Properties of  $W[\Phi,\Pi;t]$ were discussed and its equation of motion was derived. The scalar fields in thermal equilibrium and a system undergoing a phase transition due to the falling temperature were also studied. Later on  the Wigner functional of electromagnetic field was analyzed \cite{Bialynicki-Birula:2000} with the thermal, coherent and squeezed states being discussed. 

Our aim here is to introduce the Wigner functional of fermionic fields which take their values in a Grassmann algebra. While such fields are commonly used nowadays in the path-integral formulation of quantum field  theory, other applications of anticommuting numbers are much less popular. Nevertheless, both classical and quantum mechanics of systems with dynamical variables belonging to a Grassmann algebra were formulated long ago \cite{Berezin:1976eg}, and properties of phase-space densities in the two cases were briefly discussed. Later on, quasiprobability distributions of fermionic systems were analyzed in some detail in \cite{Cahill:1999}. These studies show that in spite of a different algebraic structure, fermionic objects appear to be rather similar to their bosonic counterparts. Even more, at a superficial level the corresponding formulas often look almost the same and the differences are merely in mathematical background. We encounter an analogous situation with the Wigner functional of fermionic fields which is constructed like that of bosonic ones and the properties of the two functionals are surprisingly similar. These circumstances make us adopt a specific style of presentation of our results with many -- hopefully not too many -- technical details included, as it is rather difficult to expose the differences and similarities of Wigner functionals of bosonic and fermionic fields not referring to technicalities. 

In the subsequent sections, the Wigner functional is defined and its properties as a phase-space density are discussed. The equation of motion, which has a form of the Liouville  equation, is also derived and its solutions are given. Some remarks close the paper.

\section{Definition of the Wigner functional}

Let us define the Wigner functional of fermionic field in the same way as in the case of bosonic field \cite{Mrowczynski:1994nf} that is
\begin{eqnarray}
\label{WF-fermi-def}
W[\Psi,\Pi;t] \buildrel \rm def \over = 
\int \mathcal{D}\varphi({\bf x})\ \mathrm{exp}\Big[-i\int d^3x\ \Pi({\bf x})\varphi({\bf x})\Big]
\langle \Psi({\bf x}) + \frac{1}{2}\varphi({\bf x})|\hat{\rho}(t)|\Psi({\bf x}) - \frac{1}{2}\varphi({\bf x})\rangle,
\end{eqnarray}
where $\Psi({\bf x})$ is the time-independent fermionic field in the Schr\" odinger picture, and $\int \mathcal{D}\varphi({\bf x}) \dots$ denotes here the functional Berezin integral which is extensively discussed by its founder in \cite{Berezin:1966}.

To check whether the definition (\ref{WF-fermi-def}) makes any sense, we consider for the beginning the Gaussian functional with 
\begin{equation}
\label{F11}
\langle\Psi_1({\bf x})|\hat{\rho}|\Psi_2({\bf x})\rangle=e^{-\frac{1}{2}\int d^3x [\Psi_1({\bf x})A\Psi_1({\bf x})+\Psi_2({\bf x})A\Psi_2({\bf x})]},
\end{equation}
where $\Psi_1$, $\Psi_2$ are the two-component Weyl spinors and $A$ is an antisymmetric matrix ($A^T=-A$). If the matrix $A$ is not antisymmetric, only its antisymmetric part contributes to (\ref{F11}), as squares of fermionic fields vanish. The formula (\ref{F11}) substituted into the definition  (\ref{WF-fermi-def}) provides
\begin{equation}
W[\Psi,\Pi]=e^{-\int d^3x \, \Psi ({\bf x})A\Psi({\bf x})}\int\mathcal{D}\varphi\ 
e^{-i\int d^3x \, \Pi({\bf x})\varphi({\bf x})}
e^{-\frac{1}{4}\int d^3x \, \varphi({\bf x}) A\varphi({\bf x})} .
\end{equation}
To compute the functional integral over $\varphi({\bf x})$, the field is discretized $\varphi({\bf x}) \rightarrow \varphi_1, \varphi_2, \dots \varphi_n$ and using the formula
\be
\int d\chi_m \dots d\chi_1 d\psi_m \dots d\psi_1 e^{\chi_i M_{ij} \psi_j + k_i\chi_i + p_i \psi_i}
= {\rm det}M \; e^{- k_i (M^{-1})_{ij} p_j},
\ee
which is not difficult to prove following the methods described in \cite{Berezin:1966}, the Gaussian Wigner functional is calculated as 
\begin{equation}
\label{Q6}
W[\Psi,\Pi]=C\ e^{-\int d^3x [\Psi({\bf x}) A\Psi({\bf x}) - \Pi({\bf x}) A^{-1}\Pi({\bf x})]},
\end{equation}
where $C$ is a constant. The result (\ref{Q6}), which is rather expected as fully analogous to the case of bosonic fields, suggests that the definition  (\ref{WF-fermi-def})  makes sense and is worth further consideration. 

\section{Properties of the Wigner functional}

The Wigner functional  (\ref{WF-fermi-def}) is expected to represent a density in a phase space spanned by $\Psi$ and $\Pi$. Then, there should hold the relation
\be
\label{operator-vs-Wigner}
\langle\mathcal{O}(\hat{\Psi},\hat{\Pi})\rangle
= \langle\mathcal{O}(\Psi,\Pi)\rangle,
\ee
where $\mathcal{O}(\Psi,\Pi)$ is a function of $\Psi$ and $\Pi$. We use the hats to distinguish the field operators from their `classical' counterparts. The word {\it classical} is written in the quotation marks as the anticommuting fields are not classical in the usual sense. The average of the field operators is defined as
\begin{equation}
\label{ave-operator}
\langle\mathcal{O}(\hat{\Psi},\hat{\Pi})\rangle 
 \buildrel \rm def \over = 
\frac{1}{\mathrm{Tr}[\hat{\rho}(t)]} \;
\mathrm{Tr}[\hat{\rho}(t) \mathcal{O}(\hat{\Psi},\hat{\Pi})]
\end{equation}
and that of `classical' fields in the following way:
\begin{equation}
\label{ave-Wigner}
\langle \mathcal{O}(\Psi,\Pi)\rangle 
 \buildrel \rm def \over = 
\frac{1}{Z} \;
\int\mathcal{D}\Psi \, \frac{\mathcal{D}\Pi}{2\pi}\,
\mathcal{O}(\Psi,\Pi) \, W[\Psi,\Pi;t]
\end{equation}
with
\begin{equation}
\label{Z-Wigner}
Z  \equiv \int\mathcal{D}\Psi \, \frac{\mathcal{D}\Pi}{2\pi}\, W[\Psi,\Pi;t].
\end{equation}

\subsection{Computation of $Z$}

We start with the computation of $Z$ presenting here some technical details which will be mostly skipped in the subsequent sections. Substituting the Wigner functional (\ref{WF-fermi-def}) into Eq.~(\ref{Z-Wigner}) and performing the discretization of the fields, one has
\begin{eqnarray}
\label{Z2}
Z = \int d\Psi_nd\Psi_{n-1} \dots d\Psi_1 \frac{d\Pi_n}{2\pi}
\frac{d\Pi_{n-1}}{2\pi} \dots \frac{d\Pi_1}{2\pi}
d\varphi_nd\varphi_{n-1} \dots d\varphi_1 
\exp\Big[-i\Delta V \sum_{k=1}^n\Pi_k\varphi_k\Big]
\nonumber \\ \times
\langle \Psi({\bf x}) + \frac{1}{2}\varphi({\bf x})|\hat{\rho}(t)|\Psi({\bf x}) - \frac{1}{2}\varphi({\bf x})\rangle,
\end{eqnarray}
where $\Psi_i \equiv \Psi({\bf x}_i)$, $\Pi_i \equiv \Pi({\bf x}_i)$ and  $\varphi_i \equiv \varphi({\bf x}_i)$ with $i=1,2 \dots n$; $\Delta V$ is the volume of the elementary cell coming from discretization of $\mathbb{R}^3$. The equality holds in the continuum limit which is implicitly assumed here. Now we are going to take the integrals over $\Pi_i$ which requires moving  $d\Pi_i$ right of $d\varphi_i$. If $n$ is even, the number of interchanges of Grassmann  elements is also even, and the sign of the whole expression is not altered. When $n$ is an odd number, the sign changes. One computes the integrals over $\Pi_i$ expanding the exponential and taking into account the $n-$th term which is the only one contributing to the integral. Then, one obtains 
\ba
 \nn
\int \frac{d\Pi_n}{2\pi} \frac{d\Pi_{n-1}}{2\pi} \dots \frac{d\Pi_1}{2\pi}
\exp\Big[-i\Delta V\sum_{k=1}^n\Pi_k\varphi_k\Big]
 = \frac{(-i\Delta V)^n}{n!}
 \int \frac{d\Pi_n}{2\pi} \frac{d\Pi_{n-1}}{2\pi} \dots \frac{d\Pi_1}{2\pi}
\Big[\sum_{k=1}^n\Pi_k\varphi_k\Big]^n
\\ [2mm]
\label{result-1}
= (-i\Delta V)^n 
 \int \frac{d\Pi_n}{2\pi} \frac{d\Pi_{n-1}}{2\pi} \dots \frac{d\Pi_1}{2\pi}
\Pi_1\varphi_1 \: \Pi_2\varphi_2 \dots \Pi_n\varphi_n 
= - \Big(\frac{i\Delta V}{2\pi}\Big)^n \delta^{(n)} (\varphi_1,\varphi_2, \dots \varphi_n ),
\ea
where the factor $(-1)^{n -1}$, which occurs due to interchanging $d\Pi_i$ and $\varphi_j$, is included, and the function
\be
\label{delta-phi}
\delta^{(n)} (\varphi_1,\varphi_2, \dots \varphi_n ) \equiv \; \varphi_1\varphi_2 \dots \varphi_n 
\ee
appears to be the Dirac delta-like function in the $n-$dimensional Grassmann algebra. Indeed, 
\be
\int d\varphi_nd\varphi_{n-1} \dots d\varphi_1  \; 
\delta^{(n)} (\varphi_1,\varphi_2, \dots \varphi_n ) \; 
f(\varphi_1,\varphi_2, \dots \varphi_n ) = f_0,
\ee
where $f_k$ with $k=0,1,2, \dots n$ is defined through the decomposition of an arbitrary function $ f(\varphi_1,\varphi_2, \dots \varphi_n )$ as
\be
f(\varphi_1, \varphi_2, \dots \varphi_n) = f_0 + \sum_{i=1}^n f_1(i) \:\varphi_i
+ \sum_{i_1, i_2 =1}^n f_2(i_1, i_2)\: \varphi_{i_1} \varphi_{i_2}  + \dots 
+ \sum_{i_1, i_2 ,\dots i_n =1}^n f_n(i_1, i_2, \dots, i_n) \: \varphi_{i_1} \varphi_{i_2} \dots \varphi_{i_n} .
\ee

The continuum limit of $\delta^{(n)} (\varphi_1,\varphi_2, \dots \varphi_n )$ is denoted as $\delta [\varphi]$ and 
we have the identity 
\begin{eqnarray}
\int \frac{\mathcal{D}\Pi}{2\pi} \,
\exp\Big[-i\int d^3x \, \Pi \varphi \Big] = C_\delta \delta [\varphi] ,
\end{eqnarray}
which is repeatedly used further on.  It is a complex mathematical issue to determine the constant $C_\delta$ which is, however, irrelevant for our considerations. It will be shown that the constants of this type drop out when the final average (\ref{ave-Wigner}) is computed. 

Using the result (\ref{result-1}), Eq.~(\ref{Z2}) provides
\be
\label{Z-final}
Z = C \int  \mathcal{D} \Psi \langle \Psi|\hat{\rho}(t)|\Psi  \rangle = C \, \mathrm{Tr}[\hat{\rho}(t)] ,
\ee
as
\begin{equation}
\mathrm{Tr}[\hat{\rho}(t)] 
 \buildrel \rm def \over = 
 \int \mathcal{D} \Psi \langle \Psi |\hat{\rho}(t)|\Psi \rangle .
\end{equation}

Now we are going to prove the relation (\ref{operator-vs-Wigner}) considering step by step three special cases.

\subsection{$\mathcal{O}(\hat{\Psi},\hat{\Pi})$ depends only on $\hat\Psi$}

We first consider $\mathcal{O}(\hat{\Psi},\hat{\Pi}) = \hat{\Psi}({\bf x})$. Since squares and higher powers of fermionic fields vanish, there is no other local operator made of $\hat{\Psi}({\bf x})$ except a linear function of $\hat{\Psi}({\bf x})$. 
Therefore, we compute
\begin{equation}
\langle\hat{\Psi}\rangle 
= \frac{1}{\mathrm{Tr}[\hat{\rho}]} \;
\int \mathcal{D}\Psi\langle\Psi|\hat{\rho}\hat{\Psi}|\Psi\rangle
= \frac{1}{\mathrm{Tr}[\hat{\rho}]} \;
\int\mathcal{D}\Psi \: \Psi \:\langle\Psi|\hat{\rho} |\Psi\rangle .
\end{equation}
The last equality holds because the states $|\Psi\rangle$ are by definition the eigenstates of $\hat{\Psi}$  that is $\hat{\Psi}|\Psi\rangle=\Psi|\Psi\rangle$. Computing $\langle \Psi \rangle$ exactly as $Z$, one finds
\be
\langle \Psi \rangle 
=  \frac{1}{Z} \; C 
 \int\mathcal{D}\Psi \: \Psi \:\langle\Psi|\hat{\rho} |\Psi\rangle 
= \frac{1}{\mathrm{Tr}[\hat{\rho}]} \;
\int\mathcal{D}\Psi \: \Psi \:\langle\Psi|\hat{\rho} |\Psi\rangle 
= \langle\hat{\Psi}\rangle .
\ee
As seen, the constant $C$ cancels out and consequently we will ignore constants of this type further on. If one considers an average of  a nonlocal product $\hat{\Psi}({\bf x}_1) \hat{\Psi}({\bf x}_2) \dots \hat{\Psi}({\bf x}_n)$, the result obviously is
\be
\label{product-Psi}
\langle \Psi({\bf x}_1) \Psi({\bf x}_2) \dots \Psi({\bf x}_n) \rangle 
=  \langle \hat{\Psi}({\bf x}_1) \hat{\Psi}({\bf x}_2) \dots \hat{\Psi}({\bf x}_n) \rangle .
\ee

\subsection{$\mathcal{O}(\hat{\Psi},\hat{\Pi})$ depends only on $\hat\Pi$}

As previously, we first consider $\mathcal{O}(\hat{\Psi},\hat{\Pi}) = \hat{\Pi}({\bf x})$
\begin{equation}
\label{ave-hat-Pi}
\langle\hat{\Pi}\rangle 
= \frac{1}{\mathrm{Tr}[\hat{\rho}]} \;
\int \frac{\mathcal{D}\Pi}{2\pi} \langle\Pi|\hat{\rho}\hat{\Pi}|\Pi\rangle
= \frac{1}{\mathrm{Tr}[\hat{\rho}]} \;
\int \frac{\mathcal{D}\Pi}{2\pi}  \Pi \langle\Pi|\hat{\rho}|\Pi\rangle ,
\end{equation}
where 
\be
\mathrm{Tr}[\hat{\rho}] 
 =  \int \frac{\mathcal{D} \Pi}{2\pi} \langle \Pi |\hat{\rho}|\Pi \rangle .
\ee
The same result is obviously obtained starting with the complete set of eigenstates of $\hat\Psi$. Since we will repeatedly go from the eigenstates of  $\hat\Psi$ to eigenstates of $\hat\Pi$ or {\it vice versa}, let us see how it proceeds
\ba
\langle\hat{\Pi}\rangle 
= \frac{1}{\mathrm{Tr}[\hat{\rho}]} \;
\int \mathcal{D}\Psi\langle\Psi|\hat{\rho}\hat{\Pi}|\Psi\rangle
=  \frac{1}{\mathrm{Tr}[\hat{\rho}]} \;
\int \mathcal{D}\Psi \frac{\mathcal{D}\Pi_1}{2\pi} \frac{\mathcal{D}\Pi_2}{2\pi}
\Pi_2
\langle\Psi|\Pi_1\rangle
\langle\Pi_1| \hat{\rho}|\Pi_2\rangle \langle\Pi_2|\Psi\rangle.
\ea
Because
\begin{equation}
\langle\Psi|\Pi\rangle = \exp\Big[i\int d^3x \ \Pi \Psi \Big],
\;\;\;\;\;\;\;\;\;\;\;
\langle\Pi|\Psi\rangle = \exp\Big[-i\int d^3x \ \Pi \Psi\Big],
\end{equation}
one finds
\be
\langle\hat{\Pi}\rangle 
= \frac{1}{\mathrm{Tr}[\hat{\rho}]} \;
\int \mathcal{D}\Psi \frac{\mathcal{D}\Pi_1}{2\pi} \frac{\mathcal{D}\Pi_2}{2\pi}
 \exp\Big[i\int d^3x (\Pi_1 - \Pi_2 ) \Psi \Big]
\Pi_2
\langle\Pi_1| \hat{\rho}|\Pi_2\rangle .
\ee
Now one performs the integral over $\Psi$, which generates $\delta[\Pi_1 - \Pi_2]$, and after taking the integral either over $\Pi_1$ or $\Pi_2$ one reproduces the result (\ref{ave-hat-Pi}).

Let us now compute $\langle \Pi \rangle$ as
\be
\label{ave-Pi-1}
\langle \Pi \rangle
=\frac{1}{Z}\int\mathcal{D}\Psi\ \frac{\mathcal{D}\Pi}{2\pi}\ \mathcal{D}\varphi
\exp\Big[-i\int d^3x \ \Pi \varphi \Big] \Pi
\langle \Psi + \frac{1}{2}\varphi |\hat{\rho} |\Psi - \frac{1}{2}\varphi \rangle.
\ee
With the complete sets of momentum eigenstates, the formula (\ref{ave-Pi-1}) is rewritten as
\begin{eqnarray}
\langle \Pi \rangle
= 
\frac{1}{Z} \int\mathcal{D}\Psi\frac{\mathcal{D}\Pi}{2\pi}\mathcal{D}\varphi\frac{\mathcal{D}\Pi_1}{2\pi}\frac{\mathcal{D}\Pi_2}{2\pi}
\exp\Big\{ i\int d^3x \Big[ ( \Pi_1-\Pi_2) \Psi 
+  \Big(  \frac{1}{2} (\Pi_1+\Pi_2) - \Pi \Big) \varphi \Big] \Big\}
\Pi
\langle\Pi_1|\hat{\rho} |\Pi_2\rangle.
\end{eqnarray}
The integrals over $\Psi$ and $\varphi$  produce the deltas $\delta[\Pi_1 - \Pi_2]$ and
$\delta[(\Pi_1 + \Pi_2)/2 - \Pi]$, respectively. After taking the trivial integrals over $\Pi_1$ and $\Pi_2$, one shows that 
\be
\langle\hat{\Pi}\rangle = \langle \Pi \rangle .
\ee
If one considers an average of a nonlocal product $\hat{\Pi}({\bf x}_1) \hat{\Pi}({\bf x}_2) \dots \hat{\Pi}({\bf x}_n)$, the result is obviously analogous to Eq.~(\ref{product-Psi}).

\subsection{$\mathcal{O}(\hat{\Psi},\hat{\Pi})$ depends on $\hat\Psi \hat\Pi$}

Let us compute $\langle \Pi({\bf x}) \Psi ({\bf y})\rangle$. The computation proceeds in the same way independently whether ${\bf x} = {\bf y}$ or ${\bf x} \not= {\bf y}$. So, the arguments are dropped. Using the complete sets of momentum eigenstates, we have
\ba
\label{ave-Pi-Psi1}
\langle \Pi \Psi \rangle
= \frac{1}{Z}\int \mathcal{D}\Psi \frac{\mathcal{D}\Pi_1}{2\pi} \frac{\mathcal{D}\Pi_2}{2\pi}
\exp\Big[ i\int d^3x  ( \Pi_1-\Pi_2) \Psi \Big] \
\frac{\Pi_1 + \Pi_2}{2}\: \Psi
\langle\Pi_1|\hat{\rho} |\Pi_2\rangle.
\end{eqnarray}
Now we evaluate $\langle\hat\Pi \hat\Psi \rangle$ as
\ba
\label{Pi-Psi}
\langle \hat\Pi  \hat\Psi \rangle 
=  \frac{1}{\mathrm{Tr}[\hat{\rho}]} \;
\int \mathcal{D}\Psi\langle\Psi|\hat{\rho}  \hat\Pi  \hat\Psi |\Psi\rangle
=
\frac{1}{\mathrm{Tr}[\hat{\rho}]} \;
\int \mathcal{D}\Psi  \frac{\mathcal{D}\Pi_1}{2\pi}\frac{\mathcal{D}\Pi_2}{2\pi}
\exp\Big[ i\int d^3x ( \Pi_1-\Pi_2) \Psi \Big]
\Pi_2 \Psi
\langle\Pi_1|\hat{\rho} |\Pi_2\rangle,
\ea
which is not equal to $\langle \Pi \Psi \rangle$.

In case of bosonic fields, it was found \cite{Mrowczynski:1994nf} that $\langle \Pi \Phi \rangle = \frac{1}{2} \langle \hat\Pi  \hat\Phi + \hat\Phi  \hat\Pi \rangle $; that is, the operators $\hat\Pi$ and $\hat\Phi$ have to be symmetrized. One can guess that the fermionic operators need to be antisymmertrized. To clarify this point, we compute $\langle\hat\Psi \hat\Pi \rangle$
\begin{equation}
\langle \hat\Psi  \hat\Pi \rangle 
= \frac{1}{\mathrm{Tr}[\hat{\rho}]} \;
\int \mathcal{D}\Psi\langle\Psi|\hat{\rho}  \hat\Psi  \hat\Pi |\Psi\rangle
= - \frac{1}{\mathrm{Tr}[\hat{\rho}]} \;
\int \mathcal{D}\Psi\langle\Psi|  \hat\Pi \hat{\rho}  \hat\Psi  |\Psi\rangle .
\end{equation}
The minus sign occurs because ${\rm Tr} [AB] = - {\rm Tr} [BA]$ when the elements of matrices $A$ and $B$ belong to a Grassmann algebra. Using the complete sets of momentum eigenstates,  one finds
\be
\label{Psi-Pi}
\langle \hat\Psi  \hat\Pi \rangle 
= - \frac{1}{\mathrm{Tr}[\hat{\rho}]} \;
\int \mathcal{D}\Psi  \frac{\mathcal{D}\Pi_1}{2\pi} \frac{\mathcal{D}\Pi_2}{2\pi}
\exp\Big[ i\int d^3x ( \Pi_1-\Pi_2) \Psi \Big]
\Pi_1 \Psi
\langle\Pi_1|\hat{\rho} |\Pi_2\rangle.
\ee
Combining the results (\ref{Pi-Psi}, \ref{Psi-Pi}), we prove that
 \be
\langle \Pi \Psi \rangle = \frac{1}{2} \langle \hat\Psi  \hat\Pi - \hat\Pi  \hat\Psi \rangle .
\ee
The result can be easily generalized to a nonlocal multiple product of $\hat\Psi$ and $\hat\Pi$.

We conclude this section by saying that the Wigner functional  (\ref{WF-fermi-def})  indeed represents a density in a phase space spanned by $\Psi$ and $\Phi$.

\section{Dirac field}

To proceed further, a dynamics of fermionic field must be specified. We consider here the four-component Dirac spinor $\hat\Psi(t, {\bf x})$, which is an operator acting in the Hilbert space of states,  with the Lagrangian density 
\be
\hat{\cal L}  =  \hat{\overline\Psi} \, 
\big( i \gamma^\mu \partial_\mu   - m \big) \, \hat\Psi 
+ \hat{\cal L}_I  ,
\ee
where $ \hat{\overline\Psi} \equiv \hat\Psi^\dagger \gamma^0$, $\gamma^\mu$ are the gamma matrices and $m$ is the fermion mass; the interaction term is assumed to either describe the coupling to an electromagnetic ($A^\mu$) or scalar ($\Phi$) classical field
\ba
\hat{\cal L}_I = \left\{ \begin{array}{c} 
- e \hat{\overline\Psi}  \gamma^\mu \hat\Psi \, A_\mu ,
\\
-g \hat{\overline\Psi} \hat\Psi \, \Phi 
\end{array} \right. 
\ea
with $e$ and $g$ being the coupling constants. The field operators  $\hat\Psi$ and $\hat{\overline\Psi}$ obey the Dirac equations 
\ba
\label{Dirac-eq-1}
(\gamma_\mu \partial^\mu + i \, m) \hat\Psi =
\left\{ \begin{array}{c} 
-i e \gamma^\mu A_\mu \hat\Psi,
\\
-i g \Phi \,  \hat\Psi ,
\end{array} \right.
\;\;\;\;\;\;
\hat{\overline\Psi} (\gamma_\mu \buildrel \leftarrow \over{\partial^\mu}  - i\, m)  = 
\left\{ \begin{array}{c} 
i e \hat{\overline\Psi}  \gamma^\mu A_\mu ,
\\
ig \hat{\overline\Psi} \, \Phi.
\end{array} \right.
\ea

As is well known, the momentum conjugate to $\hat\Psi$ is
\be
\hat\Pi \buildrel \rm def \over 
=   \frac{\partial \hat{\cal L}}{\partial \dot{\hat\Psi}} 
= i \hat{\overline\Psi} \gamma^0  
= i \hat\Psi^\dagger
\ee
with $\dot{\hat\Psi} \equiv \frac{\partial \hat\Psi}{\partial t}$, and the field operators obey the anticommutation relations
\be
\{ \hat{\Psi}_\alpha (t,{\bf x}), \hat{\Psi}_\beta (t,{\bf x'}) \} = 0, 
\;\;\;\;\;\;\;\;\
\{ \hat{\Pi}_\alpha (t,{\bf x}), \hat{\Pi}_\beta (t,{\bf x'}) \} = 0, 
\ee
\be
\label{com-Psi-Pi}
\{ \hat{\Psi}_\alpha (t,{\bf x}), \hat{\Pi}_\beta (t,{\bf x'}) \} = \delta^{\alpha \beta} \delta^{(3)}({\bf x} - {\bf x'}),
\ee
where $\alpha, \beta = 1,\; 2, \; 3, \; 4$ are the spinor indices.

The Hamiltonian is
\be
\label{hat-H-def}
{\hat H} \buildrel \rm def \over = \int d^3x \, \Big( \hat\Pi \dot{\hat\Psi} - \hat{\cal L} \Big)
= - \int d^3x \, \hat\Pi \gamma^0 \bigg( \boldsymbol\gamma \cdot \nabla + im 
   +  {ie \gamma^\mu A_\mu \choose  ig \Phi} \bigg) \hat\Psi 
\ee
and the Hamilton equations, which read
\ba
\label{Ham-eq-1}
\dot{\hat\Psi} &=& \frac{\delta \hat{H}}{\delta \hat\Pi} 
= - \gamma^0  
\bigg( \boldsymbol\gamma \cdot \nabla + im 
+ {ie \gamma^\mu A_\mu \choose  ig \Phi} \bigg) \hat\Psi ,
\\ [2mm]
\label{Ham-eq-2}
\dot{\hat\Pi} &=& 
-  \frac{\delta \hat{H}}{\delta \hat\Psi} 
= - \hat\Pi \gamma^0 
\bigg( \boldsymbol\gamma \cdot \buildrel \leftarrow \over \nabla -im 
-  {ie \gamma^\mu A_\mu \choose  ig \Phi} \bigg) ,
\ea
are identical with the Dirac equations (\ref{Dirac-eq-1}).

\section{Equation of motion}

The equation of motion of the Wigner functional is derived from the equation satisfied by the density matrix operator
\begin{equation}
i{\partial \over \partial t} \hat{\rho}(t) = [\hat H, \hat{\rho}(t)] ,
\label{rho-EoM}
\end{equation}
where the Hamiltonian is given by Eq.~(\ref{hat-H-def}). The field operators are assumed to be in the Schr\"odinger picture and thus the operators are time independent. We split the Hamiltonian as
\be
{\hat H} = {\hat H}_\nabla + {\hat H}_m + {\hat H}_I ,
\ee
where
\be
{\hat H}_\nabla \equiv - \int d^3x \,  \hat\Pi \gamma^0 \boldsymbol\gamma \cdot \nabla \hat\Psi ,
\;\;\;\;\;\;\;\;\;\;\;\;
{\hat H}_m \equiv - im \int d^3x \, \hat\Pi \gamma^0 \hat\Psi,
\;\;\;\;\;\;\;\;\;\;\;\;
{\hat H}_I  \equiv - i \int d^3x \, \hat\Pi  
{e \gamma^0 \gamma^\mu A_\mu \choose g \gamma^0 \Phi } \hat\Psi.
\ee
Eq.~(\ref{rho-EoM}) provides
\be
\label{EoM-0}
i{\partial \over \partial t}  W[\Psi,\Pi;t] = I_\nabla + I_m + I_I ,
\ee
where
\be
I_\lambda \equiv \int \mathcal{D}\varphi\, \mathrm{exp}\Big[-i\int d^3x \, \Pi \varphi \Big]
\langle \Psi + \frac{1}{2}\varphi |  [\hat H_\lambda, \hat{\rho}] |\Psi  - \frac{1}{2}\varphi \rangle 
\ee
with $\lambda = \nabla, m, I$. Further on we compute $I_\nabla$, $I_m$ and $I_I$.

\subsection{Mass term}

We first analyze the mass term showing some details which will be skipped in the discussion of  $I_\nabla$ and $I_I$.  The mass term has the form 
\ba
\nn
I_m 
&=&
im \int  d^3x  
\int \mathcal{D}\varphi\; 
\mathrm{exp}\Big[-i\int d^3x\, \Pi \varphi \Big]
\langle \Psi + \frac{1}{2}\varphi| 
\hat\Psi \gamma_0^T \hat\Pi \hat{\rho} 
+ \hat{\rho} \hat\Pi \gamma^0  \hat\Psi |\Psi - \frac{1}{2} \varphi \rangle  
\\ [2mm] 
\label{I-m-77}
&&
+ \; m {\rm Tr}[\gamma^0]  \int d^3x \,\delta^{(3)}({\bf x}=0)  \,  W[\Psi,\Pi;t],
\ea
where we have used the commutation relation (\ref{com-Psi-Pi}) providing
\be
\hat\Pi({\bf x}) \gamma^0 \hat\Psi({\bf x})  = - \hat\Psi ({\bf x}) \gamma_0^T \hat\Pi({\bf x}) + i {\rm Tr}[\gamma^0]  \,  \delta^{(3)}({\bf x}=0).
\ee
The last term in Eq.~(\ref{I-m-77}), which is rather pathological because of $\delta^{(3)}({\bf x}=0)$,
actually vanishes as ${\rm Tr}[\gamma^\mu]=0$. It is, however, kept here as analogous expressions in $I_\nabla$ and $I_I$ are nonzero. We will show that the pathological term from Eq.~(\ref{I-m-77})  will be exactly canceled by another term of this type. It is worth noting that such pathological terms do not show up when the equation of motion of the  Wigner functional of bosonic fields is derived \cite{Mrowczynski:1994nf}. The point is that the Hamiltonian of the scalar field is the sum of the terms which depend either on $\hat\Phi$ or $\hat\Pi$ but not on both. 

Since $\langle \Psi + \frac{1}{2}\varphi|$, $|\Psi - \frac{1}{2} \varphi \rangle $ are eigenstates of $\hat\Psi$, we have
\ba
\nn
I'_m &=& im
\int \mathcal{D}\varphi\; 
\mathrm{exp}\Big[-i\int d^3x\; \Pi \varphi \Big]
\\ [2mm] 
&\times&  \int  d^3x   
\Big[ \Big(\Psi + \frac{1}{2}\varphi \Big) \gamma_0^T
\langle \Psi + \frac{1}{2}\varphi| 
\hat\Pi \hat{\rho} 
|\Psi - \frac{1}{2} \varphi \rangle  
+
\langle \Psi + \frac{1}{2}\varphi| 
 \hat{\rho} \hat\Pi 
|\Psi - \frac{1}{2} \varphi \rangle  \gamma^0
 \Big(\Psi - \frac{1}{2}\varphi \Big) \Big] ,
\ea
where the pathological term is excluded from $I'_m$ that is
\be
\label{I-m-prime-def}
I'_m \equiv I_m - m {\rm Tr}[\gamma^0]  \int d^3x \, \delta^{(3)}({\bf x}=0) \,   W[\Psi,\Pi;t].
\ee

Now, the complete sets of momentum eigenstates are introduced and one gets
\ba
\nn
I'_m &=& im
\int \mathcal{D}\varphi\; 
\frac{\mathcal{D}\Pi_1}{2\pi} \frac{\mathcal{D}\Pi_2}{2\pi} \;
\mathrm{exp}\Big[-i\int d^3x\, \Pi \varphi \Big] 
\exp \Big\{ i\int d^3x \, \Big[ \Big( \Pi_1 - \Pi_2 \Big) \Psi +  \frac{\Pi_1 +\Pi_2}{2} \varphi\Big] \Big\}
\\ [2mm] 
&\times& 
 \int  d^3x  
\Big[ \Psi \gamma_0^T \big( \Pi_1 - \Pi_2 \big) 
+ \varphi \gamma_0^T \frac{\Pi_1 + \Pi_2}{2} \Big]  
\langle \Pi_1| \hat\rho|\Pi_2\rangle .
\ea

Observing that the identity, which is obvious for bosonic fields, holds for the fermionic ones with merely changed the sign  
\be
\label{relation-11}
i\frac{\delta}{\delta \Psi} 
\exp \Big[ i\int d^3 x \, \Pi \Psi \Big]
= \Pi \exp \Big[ i\int d^3 x \, \Pi \Psi \Big] ,
\ee
we find
\ba
\nn
I'_m &=& - m
\int \mathcal{D}\varphi\; 
\frac{\mathcal{D}\Pi_1}{2\pi} \frac{\mathcal{D}\Pi_2}{2\pi} \;
\mathrm{exp}\Big[-i\int d^3x\; \Pi \varphi \Big] 
 \\ [2mm] 
\label{I-m-11}
&\times& 
 \int  d^3x  
\Big(\Psi \gamma_0^T \frac{\delta}{\delta \Psi}  
+ \varphi \gamma_0^T \frac{\delta}{\delta \varphi}  \Big)
\exp \Big\{ i\int d^3x \; \Big[ \Big( \Pi_1 - \Pi_2 \Big) \Psi +  \frac{\Pi_1 +\Pi_2}{2} \varphi\Big] \Big\}
\langle \Pi_1| \hat\rho|\Pi_2\rangle .
\ea
The first term in Eq.~(\ref{I-m-11}) provides 
\ba
\label{Im1}
I'_{m_1} =  \int  d^3x  \; \Psi  \gamma_0^T \frac{\delta}{\delta \Psi}   W[\Psi,\Pi;t]  ,
\ea
but the second one requires further manipulations.

Using again the identity of the form (\ref{relation-11}), we get 
\ba
 \nn
I'_{m_2} & = & - i m
\int \mathcal{D}\varphi\; 
\frac{\mathcal{D}\Pi_1}{2\pi} \frac{\mathcal{D}\Pi_2}{2\pi} \;
\\ [2mm]
&\times& 
 \int  d^3x   \bigg[  \frac{\delta}{\delta \Pi} \;
\mathrm{exp}\Big[-i\int d^3x \, \Pi \varphi \Big] \bigg] \gamma_0^T
\frac{\delta}{\delta \varphi}  
\exp \Big\{ i\int d^3x \, \Big[ \Big( \Pi_1 - \Pi_2 \Big) \Psi +  \frac{\Pi_1 +\Pi_2}{2} \varphi\Big] \Big\}
\langle \Pi_1| \hat\rho|\Pi_2\rangle .
\ea
With the help of the partial integration formula, which also holds for the Berezin  integrals, we obtain
\ba
\label{Im2}
I'_{m_2} =  -  m
\int  d^3x  \,  \frac{\delta}{\delta \Pi}  \gamma_0^T \Pi \; W[\Psi,\Pi;t]   .
\ea
Since
\be
 \big(\gamma_0^T\big)_{\alpha \beta}  \frac{\delta}{\delta \Pi_\alpha ({\bf x})} \Pi_\beta ({\bf x})
= -  (\gamma_0 )_{\beta \alpha} \Pi_\beta ({\bf x}) \frac{\delta}{\delta \Pi_\alpha ({\bf x})} 
+ {\rm Tr}[\gamma_0] \delta^{(3)} ({\bf x}=0)  ,
\ee
where the spinor indices are explicitly written for clarity, we have
\ba
\label{Im2}
I'_{m_2} =  m
\int  d^3x  \,   \Pi \gamma_0 \frac{\delta}{\delta \Pi}   \; W[\Psi,\Pi;t]   
- {\rm Tr}[\gamma_0] \int d^3x \delta^{(3)} ({\bf x}=0)  \; W[\Psi,\Pi;t]  .
\ea
One sees that the pathological term encountered in Eq.~(\ref{I-m-77}) is exactly compensated by the term found in  Eq.~(\ref{Im2}). Therefore, combining the results (\ref{Im1}, \ref{Im2}), the final expression of $I_m$ reads
\be
\label{I-m-final}
I_m = -m \int  d^3x  \,  \Big[  \Psi \gamma_0^T \frac{\delta}{\delta \Psi} 
- \Pi \gamma_0 \frac{\delta}{\delta \Pi} \Big]  W[\Psi,\Pi;t]  .
\ee

\subsection{Interaction term}

The computation of the interaction term $I_I$ proceeds in exactly the same way as that of the mass term $I_m$, but $m \gamma^0$ must be replaced by either $e \gamma^0 \gamma^\mu A_\mu$ or $g\gamma^0 \Phi$. Therefore, we have 
\be
I_I =  - \int  d^3x   \bigg[ 
\Psi  
{e \gamma^0 \gamma^\mu A_\mu \choose g \gamma^0 \Phi }^T
\frac{\delta}{\delta \Psi} 
- 
\Pi 
{e \gamma^0 \gamma^\mu A_\mu \choose g \gamma^0 \Phi }\frac{\delta}
{\delta \Pi} \bigg]	   
W[\Psi,\Pi;t]  .
\ee

\subsection{Gradient term}

It is more tedious to compute the gradient term $I_\nabla$ than the mass one $I_m$, but conceptually there are no important differences. Finally, one finds
\be
I_\nabla  =
i \int  d^3x \; \Big[
\nabla \Psi \cdot (\gamma^0  \boldsymbol\gamma)^T 
\frac{\delta}{\delta \Psi}
- 
\nabla \Pi \cdot \gamma^0  \boldsymbol\gamma 
 \frac{\delta}{\delta \Pi} 
\Big]
W[\Psi,\Pi;t] ,
\ee
where the relative minus sign of the two terms results from the additional, when compared to analogous terms in $I_m$, partial integration with respect to ${\bf x}$.

\subsection{Equation of motion}

Summing up the terms $I_\nabla$, $I_m$ and $I_I$, the equation of motion of the Wigner functional (\ref{EoM-0}) gets the form
\ba
{\partial \over \partial t}  W[\Psi,\Pi;t]  &-&
\int  d^3x  \bigg[
 \Psi
\bigg(\gamma^0  \boldsymbol \gamma \cdot \nabla 
+i m\gamma^0 
+i{e \gamma^0 \gamma^\mu A_\mu \choose g \gamma^0 \Phi } \bigg)^T
\frac{\delta}{\delta \Psi}
\\[2mm] \nn
&+&
\Pi \bigg( \gamma^0  \boldsymbol \gamma \cdot \buildrel \leftarrow \over{\nabla}
-i m\gamma^0 
-i {e \gamma^0 \gamma^\mu A_\mu \choose g \gamma^0 \Phi }\bigg)
\frac{\delta}{\delta \Pi}  \bigg]
W[\Psi,\Pi;t]  = 0.
\ea
Eqs.~(\ref{Ham-eq-1},  \ref{Ham-eq-2}) show that the equation of motion can be rewritten as 
\be
\label{EoM-final}
\bigg[{\partial \over \partial t} 
+ \int d^3x  \bigg(
\frac{\delta H }{\delta \Pi_\alpha}
\frac{\delta}{\delta \Psi_\alpha}
+
\frac{\delta H}{\delta \Psi_\alpha} 
\frac{\delta}{\delta \Pi_\alpha} 
\bigg) \bigg]
W[\Psi,\Pi;t]   = 0,
\ee
where the `classical' Hamiltonian equals
\be
H = \int d^3x \bigg[ - \Pi \gamma^0 \bigg( \boldsymbol\gamma \cdot \nabla + im 
 + i {e \gamma^\mu A_\mu 
\choose
 g \Phi  }
 \bigg) \Psi 
\bigg] 
= \int d^3x \bigg[ \Pi \gamma^0 \bigg( \boldsymbol\gamma \cdot
 \buildrel \leftarrow \over{\nabla} - im 
 - i {e \gamma^\mu A_\mu 
\choose
 g \Phi  }
 \bigg) \Psi 
\bigg] .
\ee
We note that the differentiation of the `classical' Hamiltonian with respect the `classical' fields proceeds somewhat differently than the differentiation of the Hamiltonian  operator with respect to the field operators, as the `classical' fields belong to the Grassmann algebra. Therefore, the signs of $\delta \hat{H} /\delta \hat\Pi$ and $\delta H /\delta \Pi$ coincide, but those of $\delta \hat{H} /\delta \hat\Psi$ and $\delta H /\delta \Psi$ differ.

The equation of motion (\ref{EoM-final}) has a structure of the  classical Liouville equation. This is an important difference when compared to the Wigner functional of bosonic fields where the Liouville equation is obtained only when the fields are either free or  the quantum corrections to the interaction are neglected \cite{Mrowczynski:1994nf}. The fermionic field instead remains always classical except for the effects related to the Pauli principle. This is not surprising, as the Dirac equation is linear in the fermion field. 

Once the equation of motion (\ref{EoM-final}) is of the Liouville form, one immediately writes down its `elementary' solution as
\be
\label{W-solution}
W_{\rm el} [\Psi,\Pi;t] = \delta[\Psi({\bf x}) - \Psi_c(t,{\bf x})] \;  \delta[\Pi({\bf x}) - \Pi_c(t,{\bf x})] ,
\ee
where $ \Psi_c(t,{\bf x})$ and $\Pi_c(t,{\bf x})$ is the solution of `classical' equations of motion
\be
\dot\Psi_c(t,{\bf x}) = \frac{\delta H }{\delta \Pi_c(t,{\bf x})} , 
\;\;\;\;\;\;\
\dot\Pi_c(t,{\bf x}) = \frac{\delta H }{\delta \Psi_c(t,{\bf x})} ,
\ee
which coincide with the Dirac equation and its conjugate. We note that the sign in the second equation is different than usual, as the differentiation is performed in the Grassmann algebra.  The solution (\ref{W-solution}) is called `elementary' because it corresponds to a specific solution  $ \Psi_c(t,{\bf x}), \; \Pi_c(t,{\bf x})$ of the equations of motion. A general solution of Eq.~(\ref{EoM-final}) is a superposition of the elementary solutions  (\ref{W-solution})  with various $ \Psi_c(t,{\bf x}), \; \Pi_c(t,{\bf x})$.  One typically considers a solution of Liouville's equation which includes a distribution of initial values of the elementary solution.

\section{Final remarks}

Quantum mechanical problems are usually solved in terms of wave functions determined by the Schr\"odinger equation. The approach using the Wigner function, which obeys the transport-like equation of motion, offers an alternative technique. However, it has taken quite some time to develop appropriate methods \cite{Carruthers:1982fa,Kadanoff-Baym:1962,Apresyan-Kravtsov:1996,Schleich:2001}. The counterpart of the formalism employing the Wigner functional is the Schr\"odinger functional equation satisfied by the wave function which functionally depends on a field. The approach, which was first introduced for the bosonic fields \cite{Cornwall:1974vz} and then extended to fermionic ones \cite{Barnes:1978ea,Floreanini:1987gr}, has various applications. Nonperturbative studies of  gauge fields and quantum gravity can be enlisted here, see  \cite{Forkel:2005gp,Torre:2007zj} for exemplary recent developments. Hopefully, methods referring to the Wigner functionals will be also useful. 

The Wigner function is obviously not the only quasiprobability distribution which allows one for a phase-space description of quantum mechanical systems. There are of particular importance the so-called $Q$ and $P$ quasiprobability  functions which  were studied for fermion variables \cite{Cahill:1999}. Following the considerations, which are presented here, the functions can be rather easily generalized to functionals describing anticommuting fields, but it goes beyond the scope of our study.

\section*{Acknowledgments}
I am indebted to Berndt M\"uller for his suggestion to consider the Wigner functional of fermionic fields and for his comments. The collaboration with Oskar Madetko at an early stage of the project is also gratefully acknowledged. Finally I thank Alina Czajka for critical reading of the manuscript. This work was partially supported by the Polish National Science Centre under Grant No. 2011/03/B/ST2/00110.


\end{document}